\documentstyle[12pt]{article}
\begin{document}
\begin{titlepage}
\begin{center}
\today \hfill                           \hfill FERMILAB-PUB-96/147-T \\
                                        \hfill hep-ph/9609463\\
\vskip .5in

{\large \bf Naturalness Lowers the Upper Bound \\ on the Lightest Higgs boson Mass \\
in Supersymmetry\\}
\vskip 1.cm

Greg W. Anderson$^{(1)}$ \footnote {Email: anderson@fnth03.fnal.gov},
Diego J. Casta\~no$^{(2)}$ \footnote {Email: castano@fsuhep.hep.fsu.edu},
and Antonio Riotto$^{(1)}$ \footnote {Email: riotto@fnas01.fnal.gov}
\\
\vskip .25 cm
\it{$^{(1)}$
%Theoretical Physics Group \\
	    Fermi National Accelerator Laboratory \\
            P.O. Box 500, Batavia, Illinois 60510}\\
\vskip .25cm
\it{$^{(2)}$Dept. of Physics, Florida State University \\
Tallahassee, FL 32306 USA.}\\
\end{center}
%\begin{center}
%Submitted to {\it ??}
%\end{center}
\vskip .5cm
\begin{abstract}
We quantify the extent to which naturalness is lost as experimental
lower bounds on the Higgs boson mass increase, and we
compute the natural upper bound on the lightest supersymmetric Higgs 
boson mass. 
We find that it would be unnatural for the mass of the lightest
supersymmetric Higgs boson  to saturate it's maximal upper bound.  
In the absence of significant fine-tuning, the lightest
Higgs boson mass should lie below $120$ GeV, and in the most natural
cases it should be lighter than $108$ GeV.  For modest $\tan\beta$,
these bounds are significantly lower.
Our results imply that a failure to observe a light Higgs boson 
in pre-LHC experiments could provide a 
serious challenge to the principal motivation for weak-scale supersymmetry.
\end{abstract}

\end{titlepage}
\renewcommand{\thepage}{\roman{page}}
\setcounter{page}{2}
\mbox{ }

\renewcommand{\thepage}{\arabic{page}}
\setcounter{page}{1}
\setcounter{footnote}{0}
%THIS IS PAGE 1 (INSERT TEXT OF REPORT HERE)

\section{Introduction}
\def\theequation{1.\arabic{equation}}
\setcounter{equation}{0}
 
~~
The Higgs boson is the last remaining ingredient of a complete
standard model. It's persistent elusiveness is perhaps not surprising.  Within 
the framework of the standard model, there are no symmetries which can be invoked 
to make a fundamental scalar light.  The existence of a light scalar
degree of freedom which remains fundamental above the weak-scale
would argue for supersymmetry since supersymmetry provides the only explicitly known solution
to the naturalness problem which accompanies fundamental scalars~\cite{Witten}.
Of course, the Higgs boson may not be fundamental at all, and
the only testament to its existence may be the eventual unitarization of the longitudinal
$W$ scattering cross section at TeV scale energies.  However, 
although no vestige of the
Higgs boson may be seen until the LHC, a failure to observe a Higgs boson in 
pre-LHC experiments could significantly challenge the principle motivation for 
weak-scale supersymmetry, at least in its minimal forms.  

If nature is supersymmetric above the weak-scale, the allowable range of 
Higgs boson masses is considerably restricted.
In the minimal supersymmetric extension of the standard model (MSSM),
the lightest Higgs boson lies below $m_Z$ at tree level,
\begin{equation}
m_{h} \leq | \cos2\beta | m_{Z},
\end{equation} 
where $\tan\beta = v_u/v_d$ is the ratio of Higgs boson vacuum expectation values.
Quantum corrections can lift the 
light Higgs boson mass above $m_Z$ \cite{mh1}, but the magnitude of these corrections
are restricted if supersymmetry provides a successful solution to the 
naturalness problem.  Radiative corrections to the light Higgs boson mass in 
supersymmetry have been 
calculated by many authors~\cite{mh1,mh2,mh3}.
From these corrections, upper bounds for the lightest Higgs boson mass have been computed either by choosing arbitrary heavy masses for superpartners or
by demanding the theory remains perturbative up to some high scale~\cite{mh1,mh2,mh3}. 
While these upper bounds reasonably approximate an important, unexceedable 
upper-limit on the Higgs boson mass, they do not 
provide a complete picture of 
 our expectations for the mass of the 
lightest Higgs boson in supersymmetric models.  Realistically, we expect
the Higgs boson mass to be significantly lighter.
To achieve Higgs boson  masses as heavy as these upper-bounds
requires some or all superpartner masses to be much heavier than the weak-scale.
The appearance of this heavy mass scale in turn requires demonstrably 
large, unexplained cancellations among heavy masses 
in order to maintain a light weak-scale.
However, avoiding this fine-tuning is
the principle reason that supersymmetry was 
introduced at the weak-scale.

In this article, we observe that it would be quite unnatural for the lightest Higgs
boson mass to saturate the maximal upper bounds which have been previously computed.
We compute the natural upper bound on the Higgs boson masses in minimal, 
low-energy supergravity (MLES), and we show
the extent to which naturalness is lost as the experimental lower bound on the
lightest Higgs boson mass increases.
Section two provides a brief review of naturalness and how it is reliably quantified.
An analysis of the natural upper bound on the Higgs boson mass follows in section three.
We find that for $m_t < 175$ GeV,  if $m_h > 120 $ GeV, minimal low energy supergravity
does not accommodate the weak-scale naturally.  Moreover, 
in the {\it most} natural cases, $m_h < 108 $ GeV 
when $m_t < 175$ GeV.  For modest
$\tan\beta$, the natural upper-bound is even more restrictive.
In particular,
for $\tan\beta <2$ and $m_t < 175$ GeV, if $m_h > 100$ GeV large fine-tuning is
required, while the {\it most} natural values of the Higgs boson mass lie below $m_Z$.

This has important implications for challenging weak-scale supersymmetry
at collider experiments.  
In particular, if the lightest supersymmetric Higgs boson is not observed at 
CERN's $e^{+}e^{-}$ collider LEP-II, requiring 
natural electroweak symmetry breaking
in MLES will progressively increase the lower bound on
$\tan\beta$ as LEP-II increases in energy.
In the {\it most}  natural cases, if the energy of LEP-II is 
extended to $\sqrt{s}= 205 $ GeV, a light Higgs boson would be observed
provided it decays appreciably to $b \bar{b}$, but it would
not be possible to argue that natural electroweak symmetry
breaking is untenable in the minimal supersymmetric standard model
if the Higgs boson lies above the kinematic reach of LEP-II. By contrast, 
the proposed Run-III of Fermilab's Tevatron with 
${\cal L} = 10^{33} {\rm cm}^{-2}{\rm s}^{-1}$ (TeV33) can pose a 
very serious challenge to the minimal supersymmetric standard model.
The projected mass-reach for a standard model Higgs boson at
TeV33 is 100 (120) GeV with integrated luminosities of 10 (25) 
${\rm fb}^{-1}$~\cite{TeV2000}.
If the possibility that the light Higgs boson decays primarily to neutralinos
can be excluded on the basis of combined searches for superpartners
at LEP-II and the Tevatron, natural electroweak symmetry breaking 
in the minimal supersymmetric standard model will no longer be possible
if TeV33 fails to observe a light Higgs boson.
\section{Naturalness}
\def\theequation{2.\arabic{equation}}
\setcounter{equation}{0}

~~
The original and principle motivation 
for weak-scale supersymmetry is naturalness.
Supersymmetry provides the only explicitly known mechanism which allows fundamental 
scalars to be light without an unnatural fine-tuning of parameters.
Naturalness also implies
that superpartner masses can not lie much above the weak-scale if we are to avoid
the fine-tuning which would be needed to keep the weak-scale light. 
In this section, we recall the principle of naturalness and briefly review how it 
can be reliably quantified.  A more complete discussion of naturalness
criteria can be found in 
Ref.~\cite{Fine_Tuning}.
Although fine-tuning is an aesthetic criterion, once we adopt the prejudice that
large unexplained-cancellations are unnatural, a quantitative fine-tuning measure can
be constructed and placed on solid footing.  For any effective 
field theory, it is straightforward to identify whether 
large cancellations occur, and when these fine-tunings are present 
their severity can be reliably quantified.

In non-supersymmetric theories, light fundamental scalars are unnatural
because scalar particles receive quadratically divergent contributions to
their masses.  Generically, at one-loop, a scalar mass is of the form
\begin{equation}
   m_{S}^2(g) = g^2 \Lambda_{1}^2 - \Lambda_{2}^2,
\end{equation}
where $\Lambda_1 $ is the ultraviolet cutoff of the effective
theory, and $\Lambda_2$ is a bare term. 
The divergence in Eq. (2.1) 
must be almost completely cancelled against the counter term 
or the fundamental scalar will have a 
renormalized mass on the order of the cutoff.  
In supersymmetry, additional loops involving super-partners conspire to
cancel these quadratic divergences,
but when supersymmetry is broken, the cancellation is no longer
complete, and the dimensionful
terms in Eq. (2.1) are replaced by the mass splitting between
standard particles and their super-partners.

In this toy example, the cancellation is self-evident, and no 
abstract quantitative prescription is needed to determine when the 
parameters of the theory must conspire to give a light scalar mass.
We are interested in a more complicated example, and this requires
a quantitative prescription for identifying instances of
fine-tuning. In the toy example, if 
we examine the sensitivity of the scalar mass to variations 
in the coupling $g$:
\begin{equation}
\frac{ \delta m_{S}^2}{m_{S}^2} = c(m_{S}^2,g) \frac{\delta g}{g},
\end{equation}
where
\begin{equation}
   c(m_{S}^2;g) = 2\frac{g^2 \Lambda_{1}^2}{m_{S}^2(g)},
\end{equation}
the scalar mass will be unusually sensitive to minute changes in
$g$ when we arrange for large unexplained-cancellations~\cite{Wilson}:
\begin{equation}
c(m_{S}^2 \ll \Lambda^2) \gg c(m_{S}^2 \sim \Lambda^2 ).
\end{equation}
However, the bare sensitivity parameter  $c$, 
by itself is not a measure of naturalness.
Although physical quantities depend sensitively on minute variations of
the fundamental parameters when there is fine-tuning,
fine-tuning is not necessarily implied by $c\gg 1$. 
Large sensitivities can occur in a theory even
when there are no large cancellations
\footnote{For example the mass of the proton depends very sensitively on
minute variations in the value of the strong coupling constant at high
energy, but the lightness of the proton is a consequence of asymptotic freedom
and the logarithmic running of the QCD gauge coupling and 
not the result of unexplained cancellations.}. 
In particular, this is true for supersymmetric
extensions of the standard model, where it is known that bare sensitivity
provides a poor measure of fine-tuning~\cite{Fine_Tuning}.
A reliable measure of fine-tuning must compare
the sensitivity of a particular choice of parameters
$c$ to a measure of the average, global sensitivity in
parameter space, $\bar{c}$.  The naturalness measure
\begin{equation}
\gamma = c/\bar{c}
\end{equation}
will greatly exceed unity if and only if fine-tuning is
encountered~\cite{Fine_Tuning}
\footnote{Alternatively, we could define a measure of fine-tuning as the
ratio of the amount of parameter space in the theory supporting typical values of $m_S$ to
the amount of parameter space giving a unusually light value of $m_S$.  
This criterion 
is in fact equivalent to the ratio of sensitivity over typical sensitivity
\cite{Fine_Tuning}.}.
This definition is a quantitative implementation of a refined 
version of Wilson's naturalness criterion: Observable properties 
of a system should not be unusually unstable against minute variations 
of the fundamental parameters.

  In supersymmetric extensions of the standard model, as the masses of superpartners
become heavy, increasingly large fine-tuning
is required to keep the weak-scale light.  
Naturalness  places an upper bound on supersymmetry-breaking
parameters and superpartner masses.  Because the radiative
corrections to the Higgs boson mass increase with heavier superpartner
masses, naturalness translates into an upper limit on the mass 
of the lightest Higgs boson.   This limit is computed in the following section.

\section{Analysis}
\def\theequation{3.\arabic{equation}}
\setcounter{equation}{0}

~~
Following the methods of Ref. 6,
we have computed the severity of fine-tuning in the minimal
supersymmetric standard model.  For definiteness, we consider soft 
supersymmetry breaking parameters with (universal)
minimal, low-energy supergravity (MLES) boundary conditions.
We quantify the severity of large cancellations, and 
present our results as upper limits on the Higgs boson mass as a 
function of the degree of fine-tuning.  Although our quantitative results
were obtained in a framework with universal soft terms at a scale
near $10^{16}$ GeV, as motivated by MLES, we do not expect our bounds on the Higgs 
boson mass to significantly increase in models with more general
soft supersymmetry breaking masses provided they have
minimal particle content at the weak-scale.  Because there are enough
free parameters in MLES to independently adjust the parameters 
in the minimal supersymmetric standard model (MSSM) which most 
significantly increase the Higgs boson mass,
more general soft terms could allow one to increase the masses
of the squarks from the first two generations above their naturalness
limits in MLES, for example, but these new degrees of freedom
will not significantly increase the upper limit on the Higgs boson mass.
Qualitatively, our results are even more general,  if we enlarge
the particle content beyond the MSSM, the upper-limit on the lightest
Higgs boson mass can be increased~\cite{nonminimal}, but natural values of the 
lightest Higgs boson mass will lie significantly below any maximal 
upper-bounds.

Our calculation evolves the dimensionless couplings of the theory at
two-loops and includes one-loop threshold contributions and one-loop
correction to the Higgs potential.
From the resulting weak-scale parameters, we calculate
the pole masses for the Higgs bosons at one-loop
following standard diagrammatic techniques~\cite{mh2}.
The remaining next-to-leading order corrections to the Higgs boson mass arising from the 
two-loop evolution of dimensionful couplings
are small in the natural region of parameter space~\cite{mh2,mh3}.

Figures 1-3 show the naturalness of the Higgs boson mass as a function
of $\tan\beta$, $m_A$, and $m_t$, respectively.
In all three figures ideally natural solutions correspond to $\gamma =1$
and fine-tuning is implied by $\gamma \gg 1$. 
Figure 1 shows contours where the severity of
fine-tuning - $\gamma$ exceeds $2.5$, $5$, $10$ and $20$  
in the $\tan{\beta}$-$m_h$ plane for $m_t =175$ GeV.  
From Fig. 1 we see that
the mass of the lightest Higgs boson can not exceed $120$ GeV without
very significant fine-tuning, while in the most natural cases it
lies below $108$ GeV.  When $\tan\beta$ is small these limits are
even more restrictive.
Figure 2 shows naturalness contours for the lightest Higgs boson mass
in MLES as a function of the CP-odd Higgs mass, 
$m_A$ for $m_t =175$ GeV and arbitrary $\tan\beta$.  
If we restrict ourselves to modest or small values of $\tan\beta$
these curves will become more restrictive in the $m_h$ direction.
Figure 3 shows naturalness contours for the lightest Higgs boson mass
in MLES as a function of the top quark mass.  The inset in Fig. 3
displays the current uncertainty in the top quark mass, and the projected
uncertainties after run-II of Fermilab's Tevatron and 
after TeV33~\cite{TeV2000,topmass}.
Fine-tuning increases both with increasing superpartner masses and with
an increasing top quark Yukawa coupling. Therefore,
in contrast to the case of fixed superpartner masses where
the corrections to the mass squared
of the Higgs boson increases as $m^{4}_t$, for fixed
naturalness these corrections increases roughly as $m^{2}_t$.

We can assess the challenge to weak-scale
supersymmetry from Higgs boson searches at colliders 
from the natural regions of parameter space identified in
Figs.1-3. 
The dominant production mechanism for light CP-even Higgs boson
at LEP-II is Higgs-strahlung
\begin{equation}
 e^{+} e^{-} \rightarrow Z^{*} \rightarrow Z + h
\end{equation}
If Higgs boson decays into light neutralino pairs,
$h \rightarrow \tilde{\chi}^{0}_1 \tilde{\chi}^{0}_1 $,
are kinematically forbidden, $h$ will decay primarily to
$b \bar{b}$.  
An upper bound on the light Higgs mass reach in this mode 
is set by kinematics
and scales as $m_h < \sqrt{s} - m_Z -$ (a few) GeV.
The combined 95\% CL  exclusion reaches for a standard model
(SM) Higgs
boson at LEP-II are 83 (98) ((112)) GeV at $\sqrt{s} = $
175 (192) ((205)) GeV,  with integrated Luminosities of
75 (150) ((200)) ${\rm pb}^{-1}$, per experiment~\cite{LEPII}.
However, it is well known that
the observability of the lightest supersymmetric scalar $h$
can be degraded with respect to the standard model in two respects.
First, the $ZZh$ vertex carries a suppression of $\sin(\alpha - \beta)$
relative to the standard model vertex, where $\alpha$ is the mixing
angle of the CP-even Higgs scalars.
The departure of this factor from unity can be appreciable for relatively
light values of the $CP$-odd  mass $m_A$, but
it approaches one as the mass of the $CP$-odd Higgs increases.
For $m_A \mathrel{\raise.3ex\hbox {$>$}\mkern-14mu \lower0.6ex\hbox{$\sim$}}
 200$ GeV, $\cos^{2}(\beta-\alpha) < .01$. If the $CP$-odd Higgs
mass is light it may be produced and seen through 
associated production $ e^{+} e^{-} \rightarrow A \,h$,
but this mode provides
a less significant challenge to weak-scale supersymmetry because 
the CP-odd scalar mass $m_A$ is much less constrained by naturalness
arguments (see Fig. 2).  Second, 
the mass reach for the lightest Higgs $h$ can also be reduced
if $h$ decays invisibly into a pair of lightest superpartners,
$\tilde{\chi}^{0}_1 \tilde{\chi}^{0}_1$. 
This branching ratio
can approach 100\% when allowed~\cite{neutralinos},
and this mode becomes more probable as the mass of the lightest Higgs 
boson increases.
In the relatively clean environment of an $e^{+}e^{-}$ collider, a Higgs
with such invisible decays could be seen from the
acoplanar jet or lepton pair topologies resulting from the decay of the
associated $Z$, but the Higgs mass reach in this case is reduced to roughly 
half of the reach when $h$ decays visibly~\cite{LEPII}. 
When
$\sin^{2}(\alpha - \beta) BR(h\rightarrow b\bar{b})$ 
is maximal, in the {\it most} natural cases, LEP-II 
operating up to $\sqrt{s}= 205$ GeV would observe a light 
Higgs, but this energy is not large enough
to argue that natural electroweak symmetry
breaking is untenable in minimal supersymmetry if the Higgs 
boson lies above the kinematic reach of LEP-II.

Kinematically, 
the proposed Run-III of Fermilab's Tevatron with 
${\cal L} = 10^{33} {\rm cm}^{-2}{\rm s}^{-1}$ (TeV33)~\cite{TeV2000} can pose a 
very serious challenge to weak-scale supersymmetry.
The best single mode for discovery of a light Higgs boson at 
the Tevatron is $q'\bar{q} \rightarrow W h$, with 
 $h\rightarrow b \bar{b}$~\cite{SMW}.  TeV33 can probe a SM Higgs up to
100 (120) GeV with integrated luminosities of 10 (25) ${\rm fb}^{-1}$.
A Higgs boson mass in excess  of $120$ GeV would be extremely
unnatural in the MSSM. However,
the $Wb\bar{b}$ cross section from $W h$ production 
is also reduced by the
factor $BR(h\rightarrow b \bar{b}) \sin^{2}(\alpha-\beta)$. So
the significance of the challenge to weak-scale supersymmetry
from light Higgs searches at TeV33 will depend strongly
on the ability of  searches for neutralinos 
and charginos at the Tevatron and LEP-II 
to eliminate the possibility of 
$h \rightarrow \tilde{\chi}^{0}_1 \tilde{\chi}^{0}_1$, 
by raising the limits on the LSP mass. 
%for natural values of the Higgs mass.   
If this is the case, natural
electroweak symmetry breaking in the minimal supersymmetric
standard model will no longer be tenable if TeV33 achieves
$\int {\cal L}dt = 25 {\rm fb}^{-1}$
and fails to observe any signal of a Higgs boson.

%Although significantly higher masses will be explored by
%the CERN Large Hadron Collider (LHC), probing a Higgs boson mass
%in the natural range above LEP-II will be more difficult.
%Our study suggests...

\section{Conclusions}
\def\theequation{4.\arabic{equation}}
\setcounter{equation}{0}

~~
Natural choices of parameters in supersymmetric models lead to Higgs
boson masses which lie significantly below 
the maximal upper-bounds determined previously 
in the literature.
We have computed the natural upper bound on the Higgs mass in MLES,
and we have quantified the extent to which naturalness is lost
as the lower bound on $m_h$ increases.  A Higgs mass above $120$
GeV will require very large fine-tuning, while the most natural
values of the Higgs mass lie below $108$ GeV.
The natural values of the
lightest Higgs boson mass have important implications for
the challenge to weak-scale supersymmetry at colliders.
In particular, if the possibility that the Higgs decays predominantly 
to neutralino pairs can be excluded from neutralino mass limits
inferred from other superpartner searches, 
natural electroweak symmetry breaking will no longer be tenable in
the MSSM if TeV33 achieves the projected reach of $m_h = 120$ GeV
and fails to observe signals of a Higgs boson.

\section*{Acknowledgments}
GA acknowledges the support of the U.S. Department of Energy
under contract DE-AC02-76CH03000.
DC is supported by the U.S. Department of Energy under grant 
number DE-FG-05-87ER40319.  
AR is supported by
the DOE and NASA under Grant NAG5--2788.
Fermilab is operated by the Universities Research Association, Inc.,
under contract DE-AC02-76CH03000 with the U.S. Department of Energy.

\newpage
\begin{description}
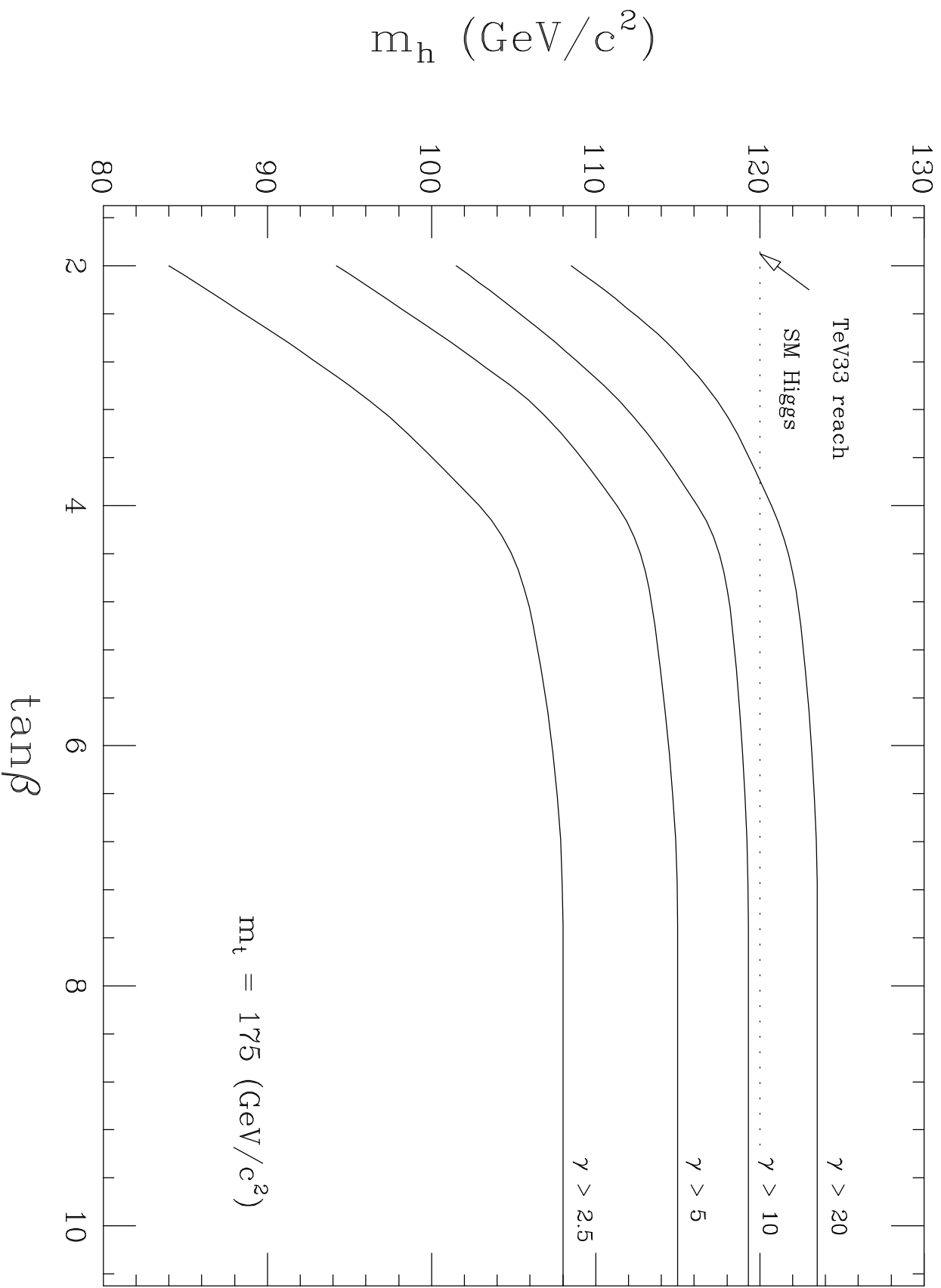
\item[\it Figure 1:] Naturalness contours for
$\gamma < 2.5, 5, 10$ and $20$ in MLES displayed 
in the $\tan\beta$ - $m_h$ plane, for $m_t = 175$ GeV. 
Ideally natural solutions correspond to
$\gamma = 1$, while fine-tuning is exhibited by $\gamma \gg 1$.

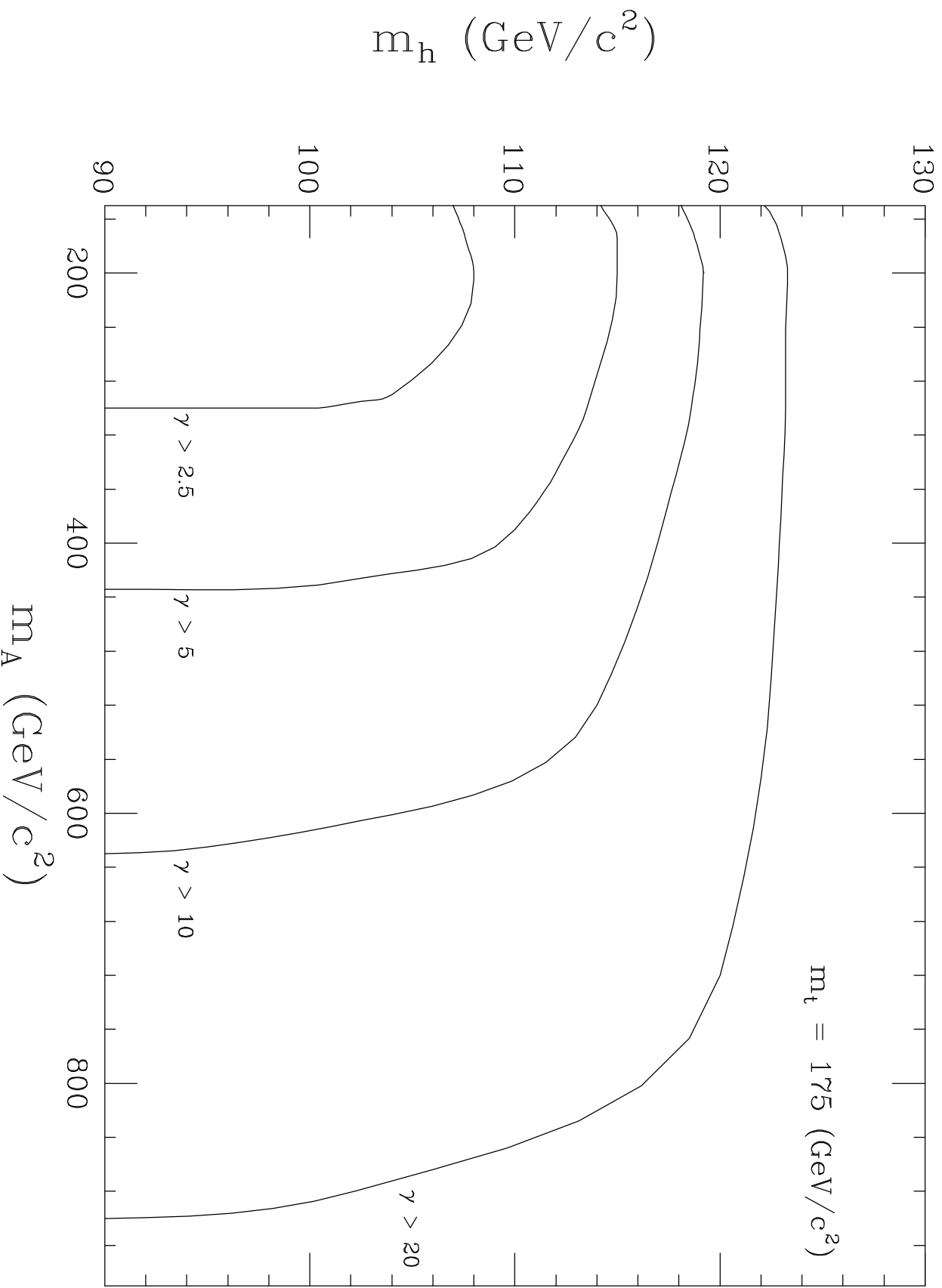
\item[\it Figure 2:] Naturalness contours for
$\gamma < 2.5, 5, 10$ and $20$ in MLES displayed 
in the $m_t$ - $m_A$ plane. More restrictive contours will
result if $\tan\beta$ is constrained to be small.  

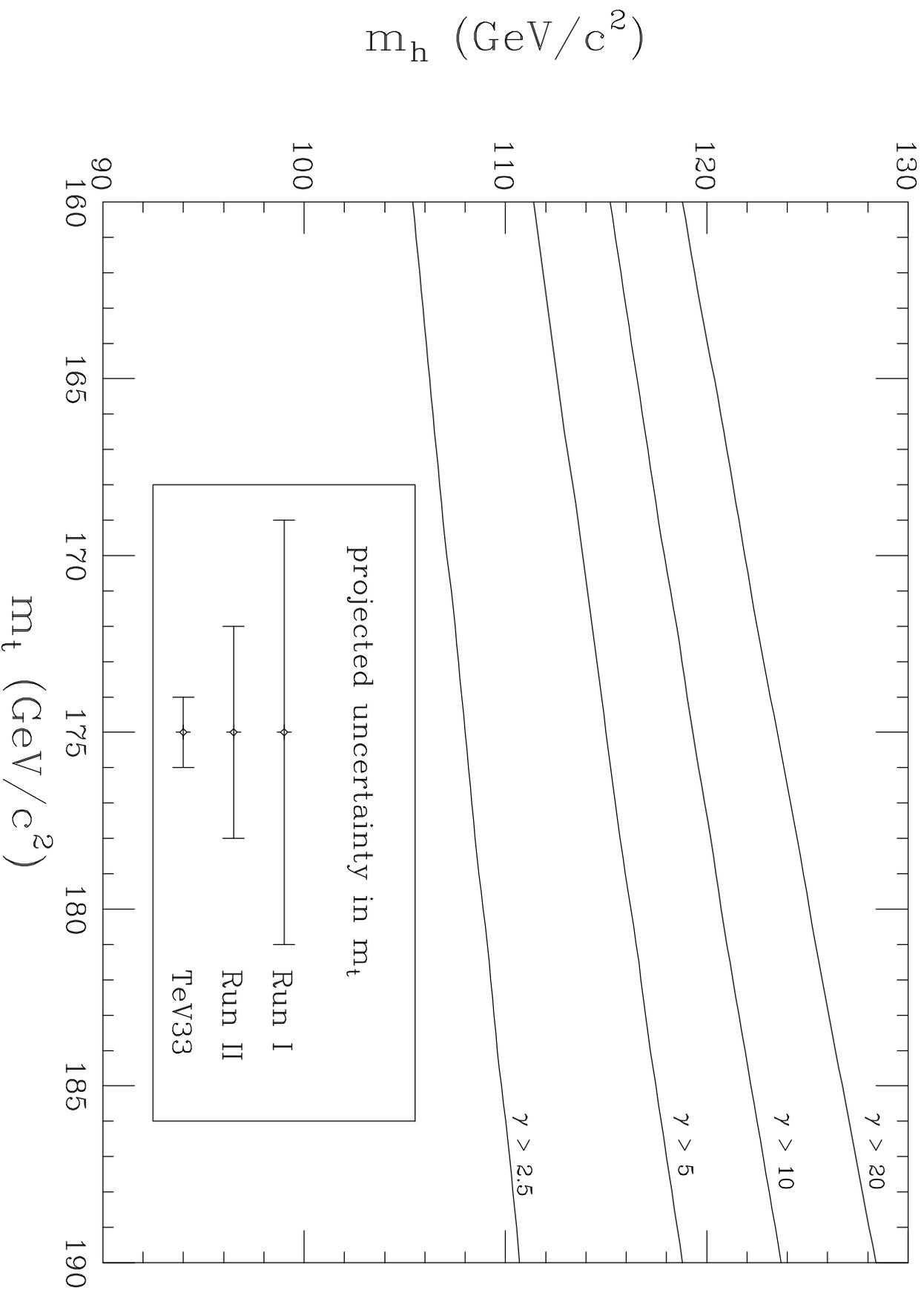
\item[\it Figure 3:] Naturalness contours for
$\gamma < 2.5, 5, 10$ and $20$ in MLES displayed 
in the $m_t$ - $m_h$ plane. More restrictive contours will
result if $\tan\beta$ is constrained to be small.  The
horizontal error bars indicate the current uncertainty in
the mass of the top quark.
\end{description}

\end{document}